\documentclass[iop,apjl,preprint,onecolumn]{emulateapj}
\usepackage{amssymb}		 
\usepackage{amsmath, amsthm, amssymb,amsfonts}
\usepackage[english]{babel}
\usepackage{epstopdf} 
\usepackage{color}
\usepackage{lipsum}% http://ctan.org/pkg/lipsum
\usepackage{color, graphicx}

\newcommand\uv{{\bf u}}
\newcommand\bbv{{\bf b}}

\newcommand\vv{{\bf v}}
\newcommand\rv{{\bf r}}

\newcommand\kv{{\bf k}}

\newcommand\mi{m_{\rm i}}
\newcommand\di{d_{\rm i}}

\newcommand\nii{n_{\rm i}}
\newcommand\nee{n_{\rm e}}
\newcommand\Pe{P_{\rm e}}
\newcommand\Bv{{\bf B}}
\newcommand\dBv{\delta{\bf B}}

\newcommand\Ev{{\bf E}}

\newcommand\Jv{{\bf J}}
\newcommand\Fv{{\bf F}}

\newcommand\Omegaci{\Omega_{\rm ci}}
\newcommand\rhoi{\rho_{\rm i}}

\newcommand\betai{\beta_{\rm i}}

\newcommand\bnabla{\boldsymbol{\nabla}}

\newcommand\Tzi{T_{0{\rm i}}}
\newcommand\Tze{T_{0{\rm e}}}

\begin{document}

\title{Subproton-scale cascades in solar wind turbulence: Driven hybrid-kinetic simulations}
\author{S.~S.~Cerri}
\email{silvio.sergio.cerri@ipp.mpg.de; silvio.cerri@df.unipi.it}
\affiliation{Max-Planck-Institut f\"ur Plasmaphysik, Boltzmannstr. 2, D-85748 Garching, Germany}
\affiliation{Physics Department ``E. Fermi'', University of Pisa, Largo B. Pontecorvo 3, 56127 Pisa, Italy}
\author{F.~Califano}
\affiliation{Physics Department ``E. Fermi'', University of Pisa, Largo B. Pontecorvo 3, 56127 Pisa, Italy}
\author{F.~Jenko}
\author{D.~Told}
\affiliation{Department of Physics and Astronomy, University of California, Los Angeles, CA 90095, USA}
\author{F.~Rincon}
\affiliation{Universit\'e de Toulouse; UPS-OMP; IRAP; 14 avenue Edouard Belin, F-31400 Toulouse, France}
\affiliation{CNRS; IRAP; 14 avenue Edouard Belin, F-31400 Toulouse, France}

%**************%
%   ABSTRACT   %
%**************%
\begin{abstract}

A long-lasting debate in space plasma physics concerns the nature of subproton-scale fluctuations in solar wind (SW) turbulence. Over the past decade, a series of theoretical and observational studies were presented in favor of either kinetic Alfv\'en wave (KAW) or whistler turbulence. Here, we investigate numerically the nature of the subproton-scale turbulent cascade for typical SW parameters by means of unprecedented high-resolution simulations of forced hybrid-kinetic turbulence in two real-space and three velocity-space dimensions. Our analysis suggests that small-scale turbulence in this
model is dominated by KAWs at $\beta\gtrsim1$ and by magnetosonic/whistler fluctuations at lower $\beta$. The spectral properties of the turbulence appear to be in good agreement with theoretical predictions. A tentative interpretation of this result in terms of relative changes in the damping rates of the different waves is also presented. Overall, the results raise interesting new questions about the properties and variability of subproton-scale turbulence in the SW, including its possible dependence on the plasma $\beta$, and call for detailed and extensive parametric explorations of driven kinetic turbulence in three dimensions.

\end{abstract}

%\pacs{52.25.Xz, 52.25.Dg, 95.30.Qd}

\maketitle

%******************%
%   INTRODUCTION   %
%******************%
\section{Introduction}\label{sec:intro}

The solar wind (SW) plasma, an ideal laboratory for the study of collisionless plasma dynamics, is mostly found in a turbulent state~\citep{BrunoCarboneLRSP2013}. Subproton-scale (``dissipation range'') turbulence in the SW has become a major research topic over the past decade, both for in-situ satellite measurements~\citep{BalePRL2005,AlexandrovaAPJ2008,AlexandrovaPRL2009,%
AlexandrovaSSRv2013,SahraouiPRL2009,SahraouiPRL2010,ChenPRL2010,ChenPRL2013,NaritaGRL2011,%
HeAPJL2012,RobertsAPJ2013} and for numerical~\citep{ShaikhMNRAS2009,ShaikhMNRAS2009b,%
ValentiniPRL2010,HowesPRL2011,ServidioPRL2012,ServidioAPJL2014,PassotEPJD2014,%
FranciAPJ2015,FranciAPJL2015,ToldPRL2015} and theoretical~\citep{StawickiJGR2001,%
GaltierPOP2003,HowesJGR2008,SchekochihinAPJS2009,GaryJGR2009,MithaiwalaPOP2012,%
BoldyrevAPJL2012,BoldyrevAPJ2013,BoldyrevAPJ2015} studies. Spacecraft observations provide important constraints on turbulent spectra, revealing the presence of breaks in the electromagnetic fluctuations around the proton kinetic scales~\citep{BalePRL2005,AlexandrovaAPJ2008,AlexandrovaPRL2009,SahraouiPRL2009,SahraouiPRL2010,ChenPRL2010}. At subproton scales, typical slopes for the magnetic energy spectrum are found to be in the range $[-2.5, -3]$, while preliminary results about its electric counterparts are in the range $[-0.3, -1.3]$. From a theoretical point of view, possible explanations for the observed spectra are the development of a kinetic Alfv\'en wave (KAW) cascade and/or a whistler cascade~\citep{BalePRL2005,SahraouiPRL2010,NaritaGRL2011,HeAPJL2012,RobertsAPJ2013,ChenPRL2013,StawickiJGR2001,GaltierPOP2003,%
HowesJGR2008,SchekochihinAPJS2009,GaryJGR2009,MithaiwalaPOP2012,BoldyrevAPJL2012,BoldyrevAPJ2013,BoldyrevAPJ2015}. 
However, the predicted energy spectra are the same for the two cases and thus auxiliary methods have been suggested in order to identify the exact nature of turbulent fluctuations~\citep{BalePRL2005,SahraouiPRL2010,NaritaGRL2011,HeAPJL2012,RobertsAPJ2013,ChenPRL2013}. Observational evidence points towards a KAW-dominated scenario for a $\beta\sim1$ plasma~\citep{SahraouiPRL2010,HeAPJL2012,RobertsAPJ2013,ChenPRL2013} ($\beta$ is the ratio between the thermal and the magnetic pressures), although contradictory results have also been reported~\citep{NaritaGRL2011}. Theoretical studies, on the other hand, have suggested that oblique KAWs and whistlers could coexist as the plasma parameters vary in space and time~\citep{StawickiJGR2001,GaryJGR2009,MithaiwalaPOP2012}. 
So far, numerical simulations have focused only on one scenario at a time, not on a possible coexistence or a transition between those cascades, leaving such a question as an open problem in SW turbulence research.

In this Letter, we wish to tackle the fundamental question of a possible dependence of the physics of subproton-scale kinetic turbulence on the plasma $\beta$ parameter by carrying out high-resolution 2D3V simulations of forced plasma turbulence as described by a hybrid Vlasov--Maxwell (HVM) model with fluid electrons. While not retaining electron kinetic effects, this approach allows for both KAWs and whistlers to be present, and it fully captures the ion kinetic physics. Besides, this 2D3V setting allows us to include large ``fluid'' scales while still fully resolving subproton scales, which is not currently possible in 3D3V due to computational limitations. Due to the intrinsic anisotropy of the turbulent MHD cascade, and to the strong damping of the parallel modes via resonances~\citep{HowesJGR2008,SchekochihinAPJS2009,GaryJGR2009,HeAPJL2012}, we also expect such ``2.5D'' simulations to retain some important dynamical features of the fully 3D case.

\section{The model}

In the HVM model, fully kinetic ions are coupled with massless fluid electrons~\citep{MangeneyJCP2002,ValentiniJCP2007,ServidioJPP2015}. The HVM equations normalized with respect to the ion mass $\mi$, the ion gyrofrequency $\Omegaci$, the Alfv\'en speed $v_A$ and the ion skin depth $\di=v_{\rm A}/\Omegaci$ are given by
%%%%%%%%%%%%%%%%%%%%%%%%%%%%%%%%%%%%%%%
\begin{eqnarray}
&& \partial_t f +
 \vv\cdot\bnabla f +
 (\Ev+\vv\times\Bv+\Fv)\cdot\bnabla_{\vv}f 
 =0\,,\label{eq:HVM_Vlasov}\\
&& \Ev \,=\, -\,\uv\times\Bv \,+\,
 \Jv\times\Bv/n \,-\,
 \bnabla\Pe/n\,, \label{eq:HVM_Ohm}\\
&& \partial_t\Bv= -\bnabla\times\Ev\,,\quad
  \bnabla\times\Bv = \Jv\,, \label{eq:HVM_Maxwell}
\end{eqnarray}
%%%%%%%%%%%%%%%%%%%%%%%%%%%%%%%%%%%%%%%
where $f=f(\rv,\vv,t)$ is the ion distribution function, $\Ev$ and $\Bv$ are the electric and magnetic fields, respectively, and $\Jv$ is the current density. We assume quasi-neutrality
$\nii\simeq\nee=n$. The number density $n$ and the ion mean velocity $\uv$ are computed as the velocity moments of $f$. An isothermal equation of state is assumed for the scalar electron pressure $\Pe$, with a given initial electron-to-ion temperature ratio $\tau=\Tze/\Tzi$.
$\Fv(\rv,t)$ is a $\delta$-correlated in time, external forcing that injects momentum in the system with a prescribed average power density $\varepsilon$. Its correlation tensor in Fourier space reads $\langle F_{\kv,i}(t)F_{\kv,j}^*(t')\rangle=\chi(k)\delta(t-t')\left[\alpha_1\left(1-k_ik_j/k^2\right)+\alpha_2\left(k_ik_j/k^2\right)\right]$, where brackets denote ensemble averaging, $\kv$ is a wave vector, $\chi(k)$ is a scalar function depending on the amplitude of the wavenumber only, and $\alpha_1$ and $\alpha_2$, respectively, quantify the relative  degrees of incompressibility and compressibility of the forcing. In all simulations presented in this Letter, we use $\alpha_1=\alpha_2=1/2$. While it may overestimate the actual compressible component of the driving in the SW context, this choice can be justified by the lack of scale separation in the simulations between the driving and ion scales, at which a mixture of compressible and incompressible fluctuations is found in the solar wind~\citep[see, e.g.,][]{AlexandrovaAPJ2008,AlexandrovaSSRv2013,KiyaniApJ2013}, and by the desire to not artificially direct energy into a particular mode at large scale. Our numerical implementation of this forcing is a direct transposition of a widely used hydrodynamic technique~\citep{AlveliusPOF1999}.

\subsection{Simulation setup}

Equations (\ref{eq:HVM_Vlasov})-(\ref{eq:HVM_Maxwell}) are solved in a 2D3V phase space using an Eulerian algorithm~\citep{MangeneyJCP2002,ValentiniJCP2007}, with fully three-dimensional vector fields. 
The initial condition is a Maxwellian plasma in a constant perpendicular magnetic field $B_0=1$. The system is initially perturbed by random, 3D, large-scale, small-amplitude magnetic fluctuations, $|\dBv(\rv)|\ll B_0$ (with wavenumbers $k_\perp\equiv(k_x^2+k_y^2)^{1/2}$ in the range $0.1\leq
k_\perp\di\leq0.3$). The driving procedure and amplitude (in code units) is identical for all the cases documented below. The average power input of $\Fv$ is $\varepsilon=5\times10^{-4}$ and the forcing acts on the smallest wave numbers of the system, $0.1\leq k_{\perp,{\rm F}}\di\leq0.2$, thus injecting energy only at the largest scales admitted by our numerical box. In the following, we consider three different initial plasma beta values ($\betai=0.2$, $1$ and $5$) and a temperature ratio $\tau=1$, i.e., typical of SW parameters. We use $1024^2$ uniformly distributed grid points to discretize a squared simulation box with $L = 20\,\pi\,\di$, corresponding to a resolution $\Delta\ell \simeq 0.06\,\di$. Doubly periodic boundary conditions are imposed, and the spectral domain spans a perpendicular wavenumber range $0.1\leq k_\perp\di\leq51.2$. Spectral filters~\citep{LeleJCP1992} on the electromagnetic fields are applied during the simulation, in order to avoid spurious numerical effects at the smallest scales: this determines the cutoff in the energy spectra at $k_\perp\di>10$. The velocity domain is limited by $v_{\rm max} = \pm\,5\,v_{{\rm th,i}}$ in each
$v$-direction, with $51^3$ uniformly distributed grid points, so $\Delta v = 0.2\,v_{{\rm th,i}}$. The time step is constrained by the CFL conditions~\citep{MangeneyJCP2002}.

\section{Hybrid-kinetic turbulence}

We first investigate the spectral properties of the statistically quasi-steady turbulent state and whether they reproduce the phenomenology expected for KAWs or whistlers. The analysis is performed at about $\sim\,11\,\tau_{\rm NL}(L)$, where $\tau_{\rm NL}(L)\sim L^{2/3}/\varepsilon^{1/3}$ is  the outer-scale nonlinear time (estimated from a Kolmogorov argument). In this regime, the average modulus of the in-plane magnetic field, $B_\perp=(B_x^2+B_y^2)^{1/2}$, remains relatively low, $\langle B_\perp/B_0\rangle\lesssim 0.08$. Nevertheless, larger values (up to $B_\perp\simeq0.5$) are observed locally in space and time, and coherent magnetic structures are formed. On the one hand,
non-negligible in-plane magnetic fluctuations allow for finite $k_\|\equiv\kv\cdot\bbv=k_xb_x+k_yb_y$ ($\bbv=\Bv/|\Bv|$ is the local unit vector along $\Bv$), i.e., for parallel kinetic effects and
oblique waves with non-zero $k_\|$. On the other hand, the local in-plane magnetic field turns out to be randomly oriented in the fully turbulent regime, and the spectra are globally isotropic in the $(k_x,k_y)$-plane. Therefore a shell-averaging technique can be adopted in that spectral plane ($k_\perp$-reduction), without being polluted by any preferential direction. Moreover, spectra are time-averaged over about $15\,\Omegaci^{-1}$. 

\begin{figure}[!t]
 \begin{minipage}[!h]{0.495\textwidth}
  \flushleft\includegraphics[width=1.05\textwidth]{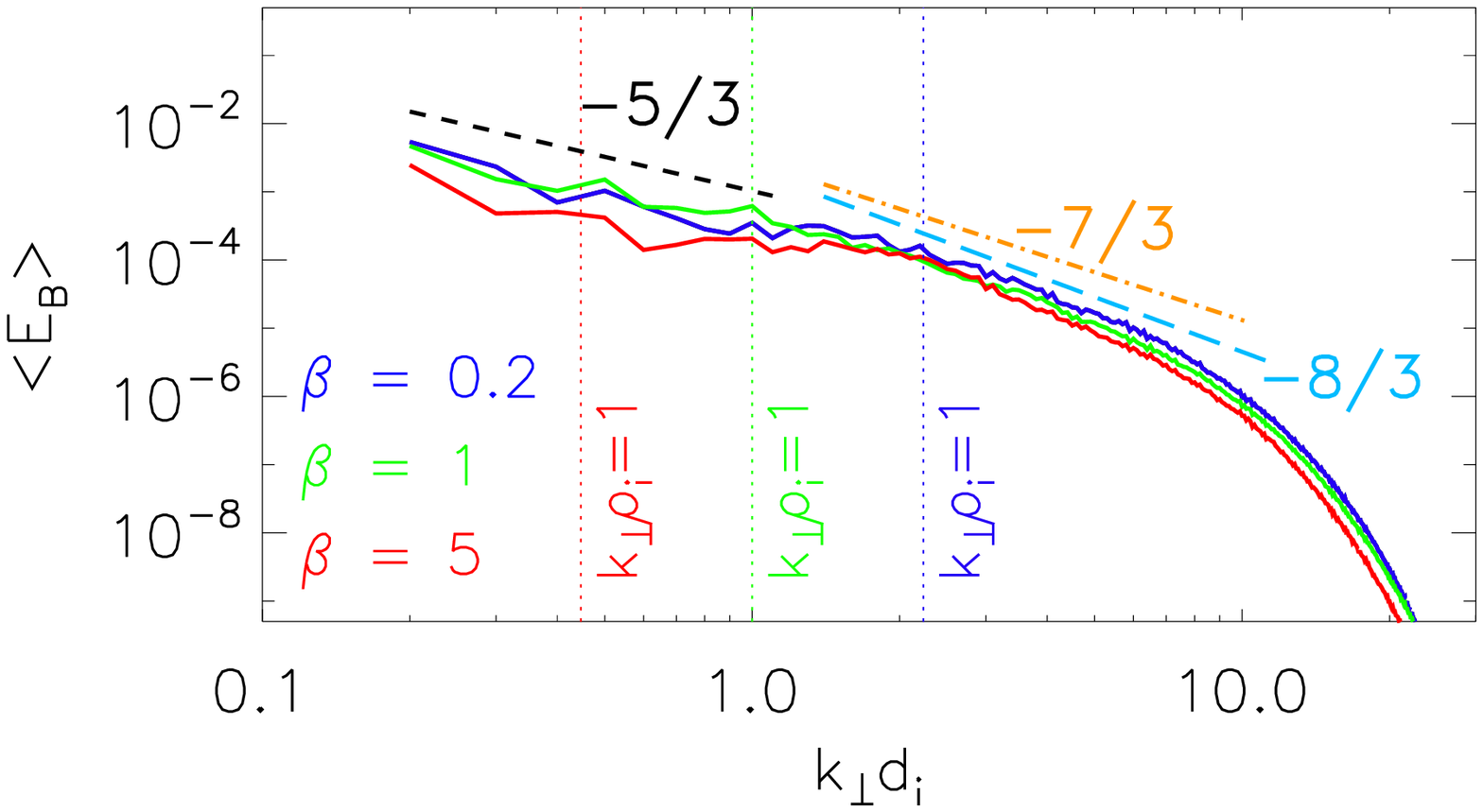}%\\
 \end{minipage} 
 \begin{minipage}[!h]{0.495\textwidth}
  \flushleft\includegraphics[width=1.05\textwidth]{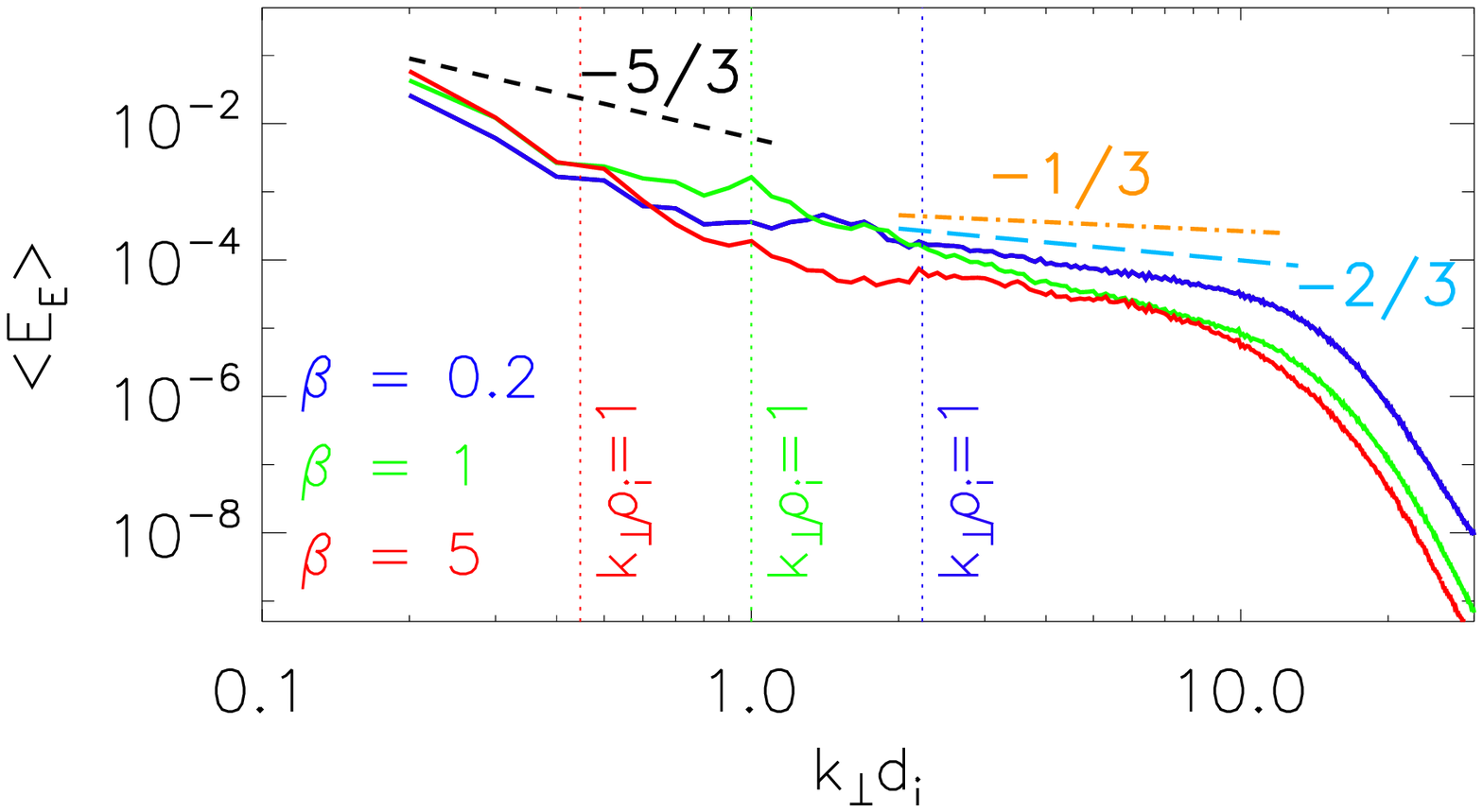}
 \end{minipage} 
 \caption{Time-averaged magnetic and electric energy spectra, $E_B(k_\perp)$ and $E_E(k_\perp)$ (left and right  panel, respectively), for  $\betai=0.2, 1, 5$, blue, green and red color (grey scale), respectively.}
 \label{fig:Spectra_comparison}
\end{figure}

In Fig.~\ref{fig:Spectra_comparison} we plot the total magnetic and electric energy spectra, $E_B(k_\perp)$ and $E_E(k_\perp)$. The $E_B$ spectrum at $k_\perp\rhoi<1$ exhibits a slope close to $-5/3$, although this result should be treated with caution because of the vicinity of the injection scale and of the small extent of the range (especially at $\betai=5$).

%================================
\begin{figure}[!t]
 \begin{minipage}[!h]{0.325\textwidth}
  \flushleft\includegraphics[width=1.125\textwidth]{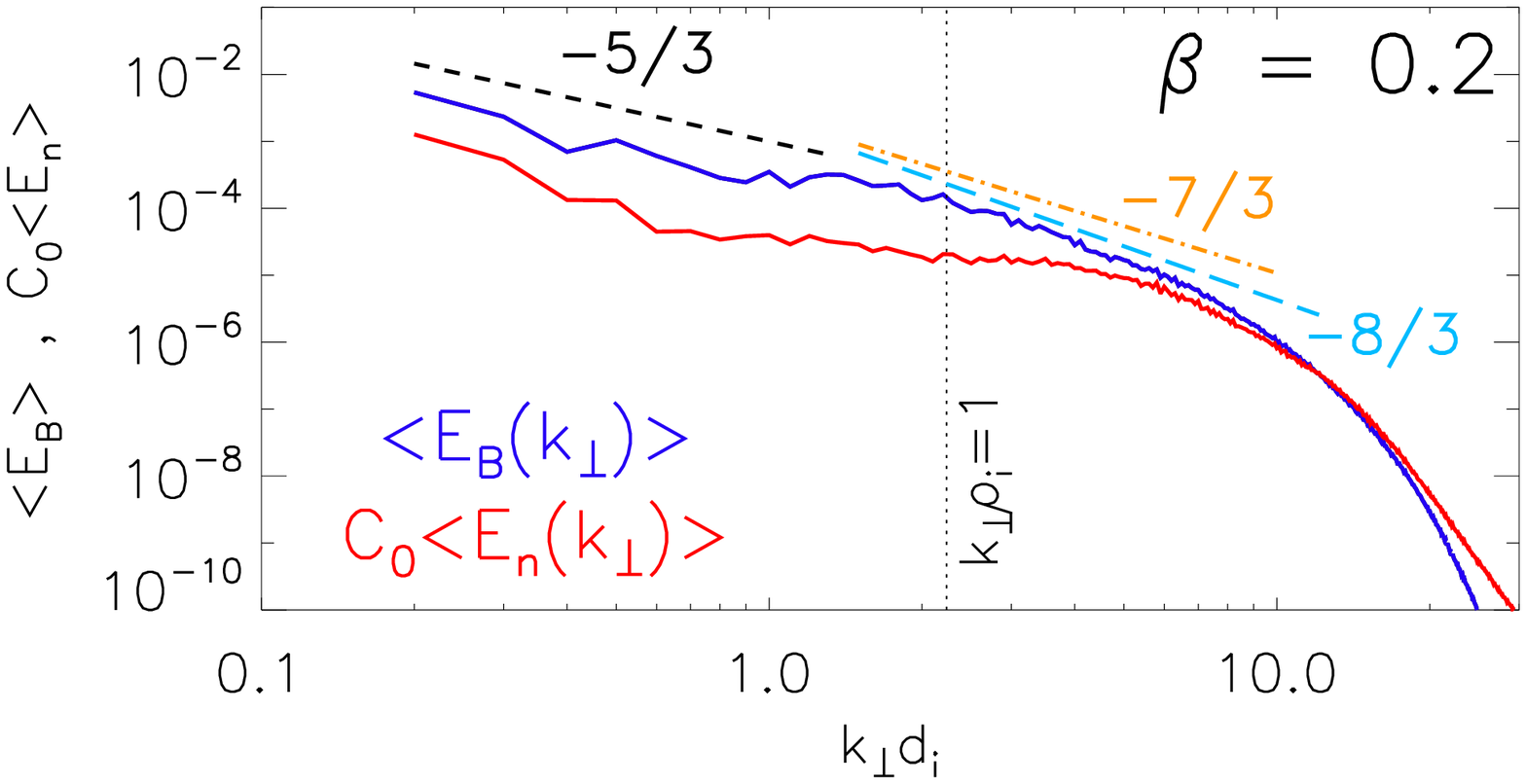}
 \end{minipage}
 \begin{minipage}[!h]{0.325\textwidth}
  \flushleft\includegraphics[width=1.125\textwidth]{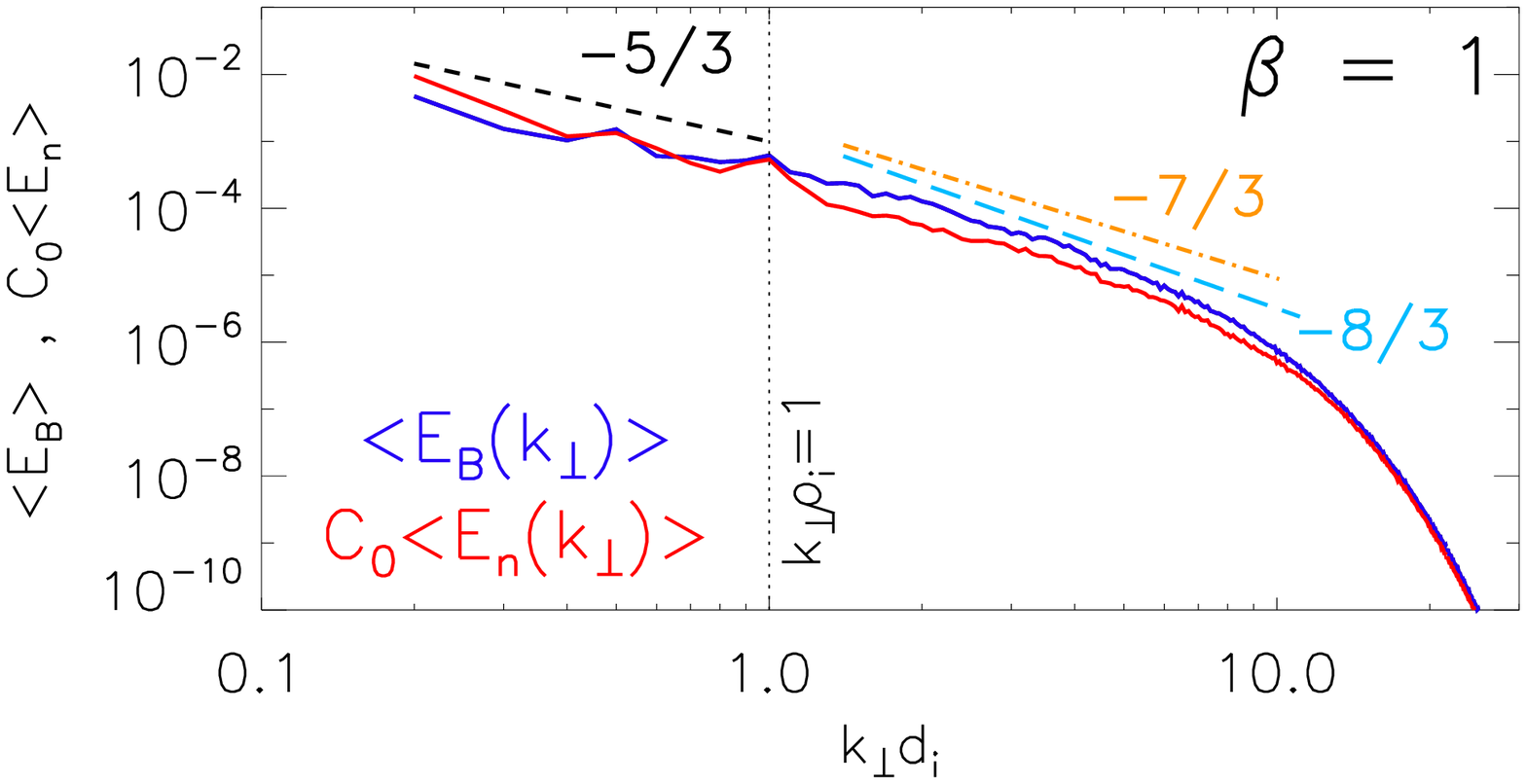}
 \end{minipage}
 \begin{minipage}[!h]{0.325\textwidth}
  \flushleft\includegraphics[width=1.125\textwidth]{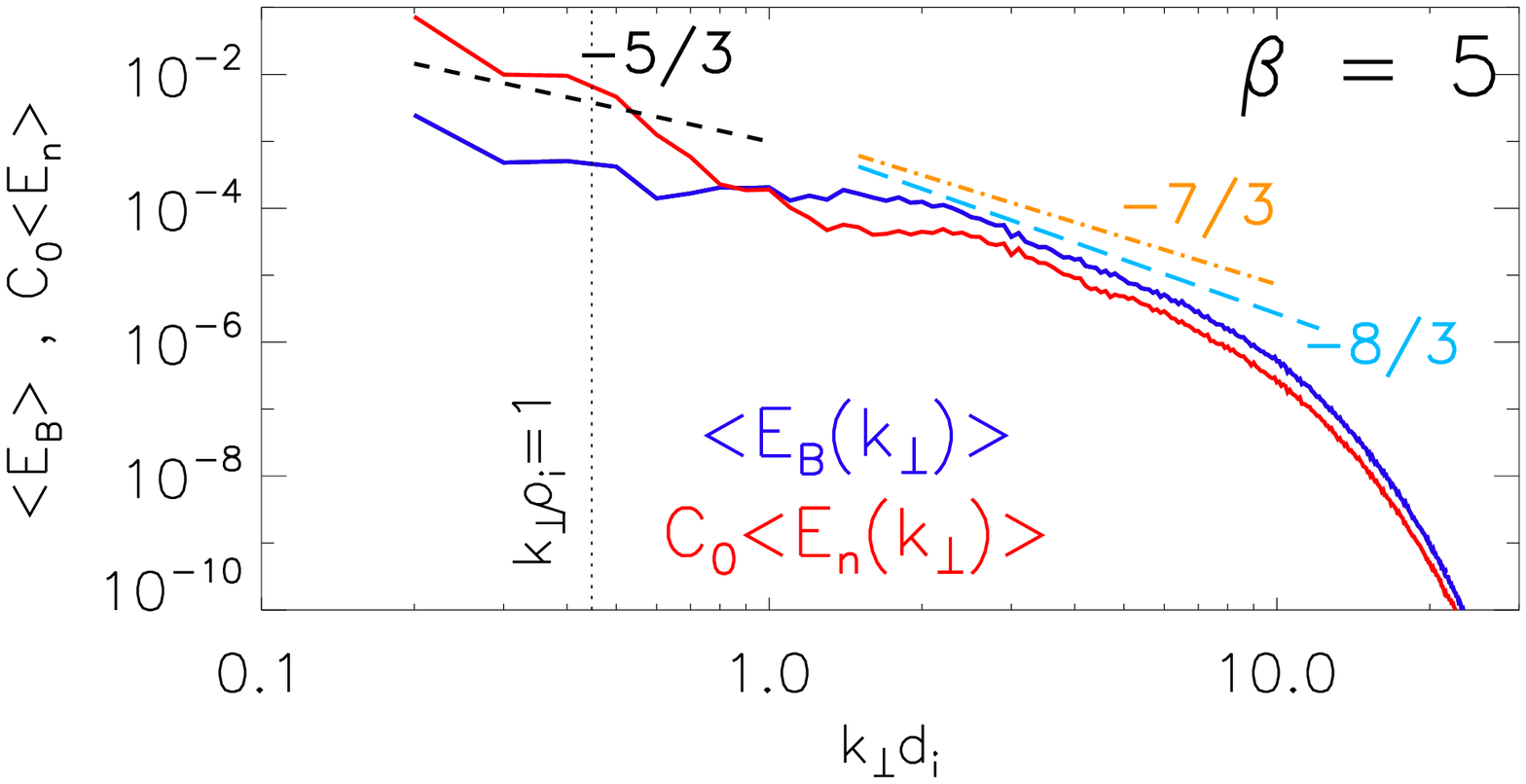}
 \end{minipage}
 \caption{Time-averaged magnetic and (normalized) density spectra (blue/black and red/gray, respectively).}
 \label{fig:Spectra_KAW-or-whistler_A}
\end{figure}

\begin{figure}[!h]
 \begin{minipage}[!h]{0.325\textwidth}
  \flushleft\includegraphics[width=1.125\textwidth]{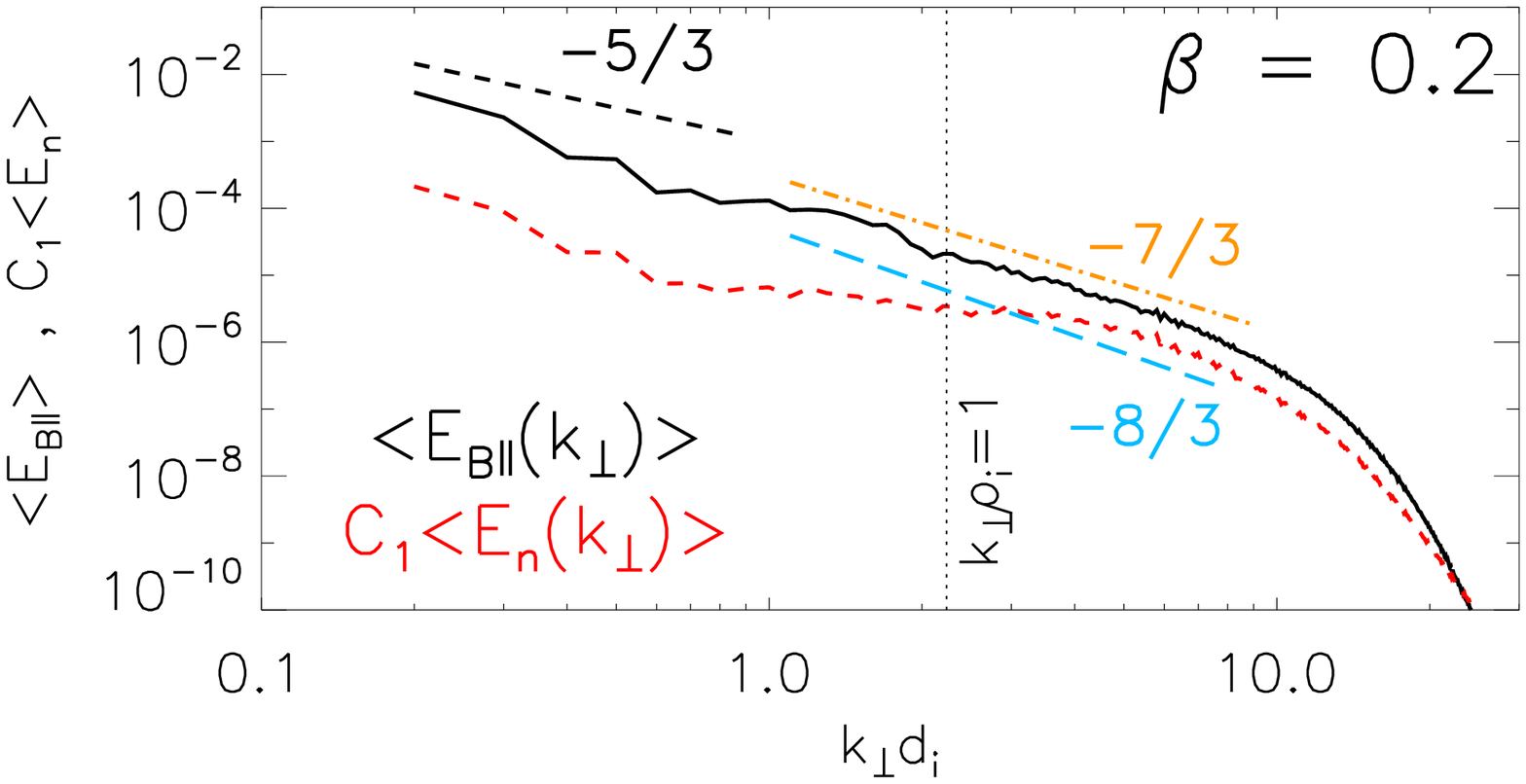}
 \end{minipage}
 \begin{minipage}[!h]{0.325\textwidth}
  \flushleft\includegraphics[width=1.125\textwidth]{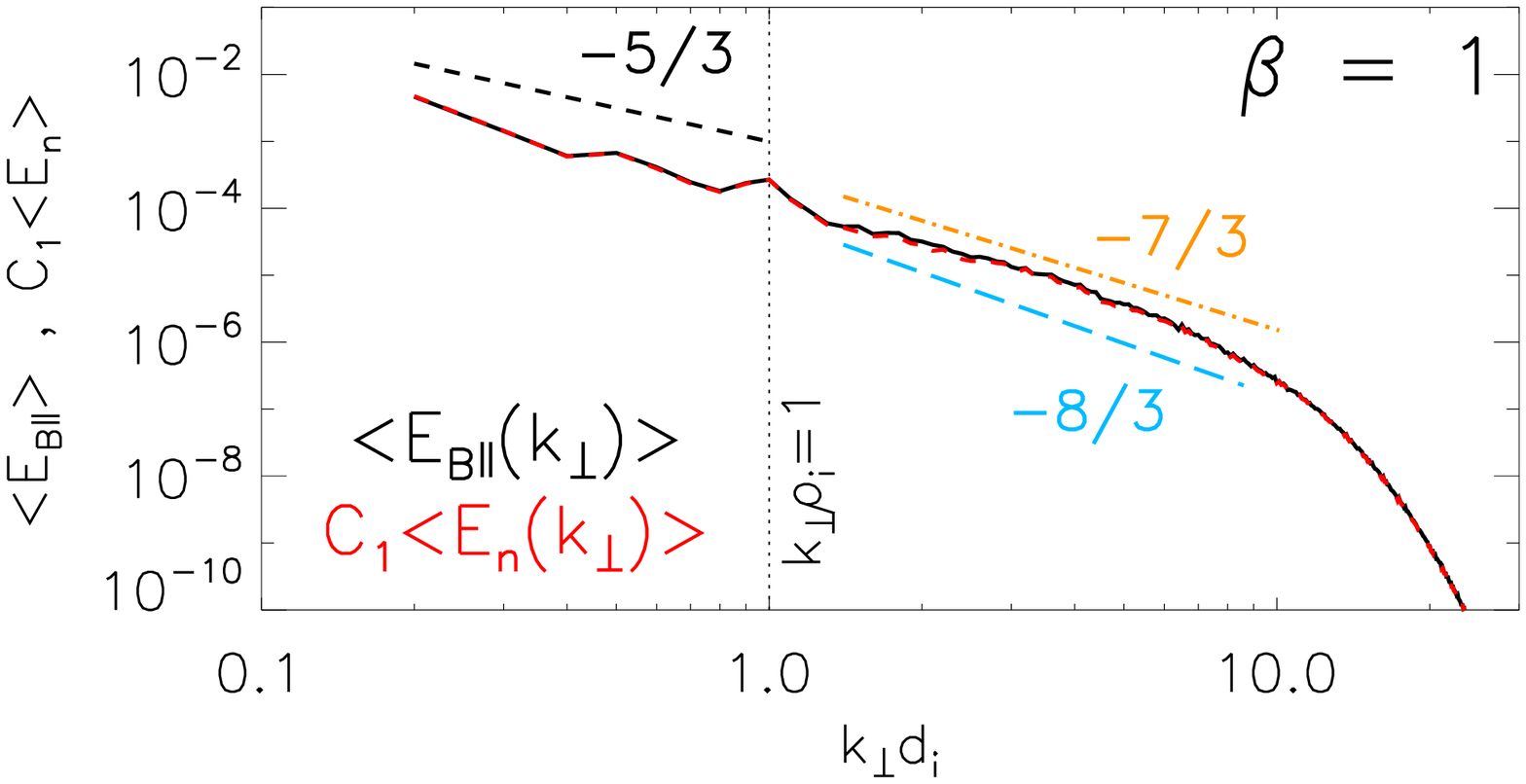}
 \end{minipage}
 \begin{minipage}[!h]{0.325\textwidth}
  \flushleft\includegraphics[width=1.125\textwidth]{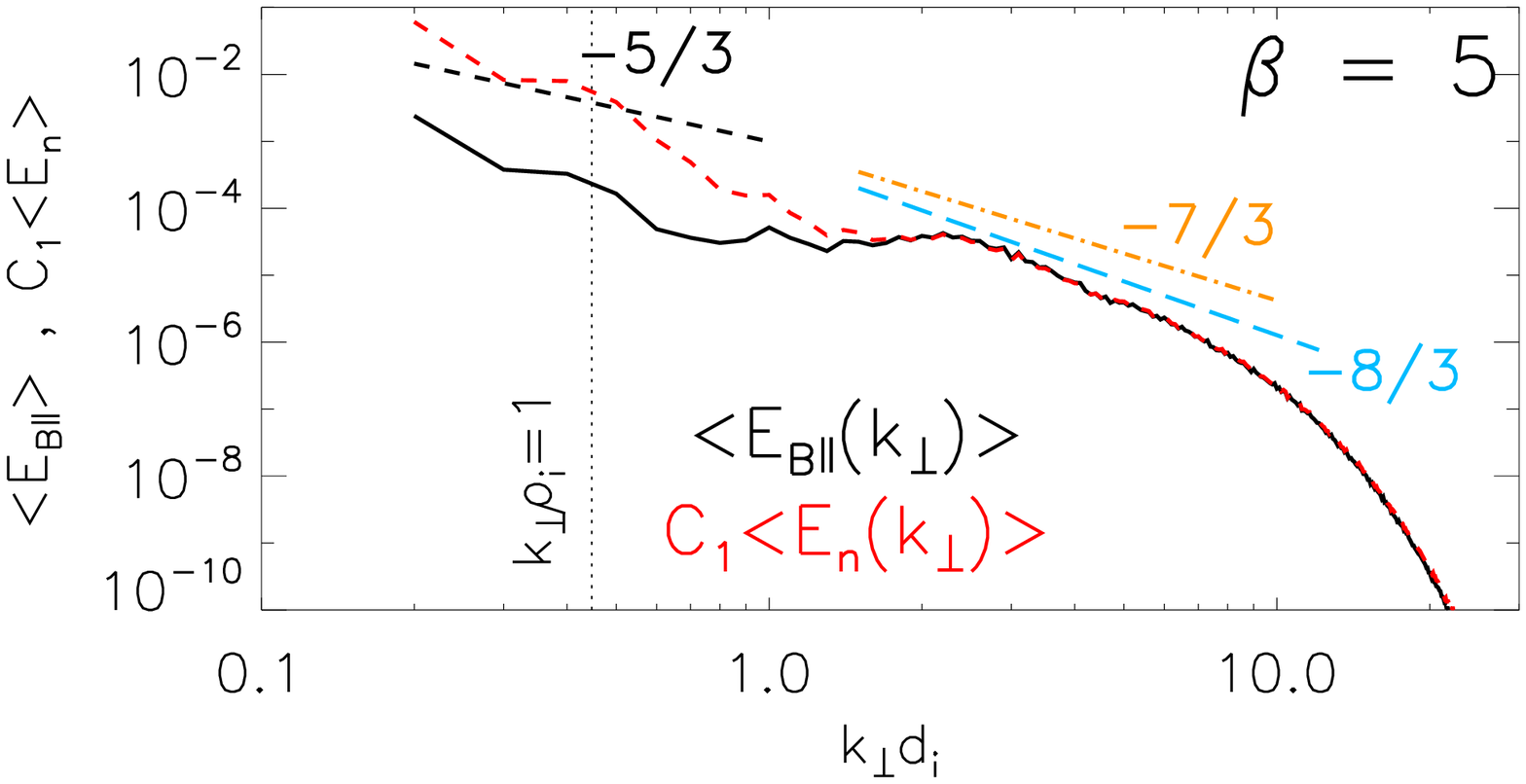}
 \end{minipage}
 \caption{Time-averaged parallel magnetic and (normalized) density spectra (black and red/grey, respectively).}
 \label{fig:Spectra_KAW-or-whistler_B}
\end{figure}

%================================ 
At $k_\perp\di>1$, the spectral index changes for all three cases and lies between $-8/3$ and $-3$, in general agreement with spacecraft observations~\citep{BalePRL2005,AlexandrovaAPJ2008,SahraouiPRL2009,AlexandrovaPRL2009,ChenPRL2010,SahraouiPRL2010}. 
On the one hand, at $\betai=0.2$ and $1$, the $E_{B\perp}$ (not shown) and $E_B$ spectra for $k_\perp\di>1$ appear to be fitted better with a $-8/3$ slope (Fig.~\ref{fig:Spectra_KAW-or-whistler_A}), while the $E_{B\|}$ (and $E_n$, at $\betai=1$) spectrum is well fitted by a $-7/3$ slope (Fig.~\ref{fig:Spectra_KAW-or-whistler_B}). A $-8/3$ slope would be in agreement with theory for fluctuations forming two-dimensional structures~\citep[][; coherent structures are indeed visible in our simulations]{BoldyrevAPJL2012}, whereas $-7/3$ is the prediction of the standard theories of KAW and whistler turbulence~\citep{GaltierPOP2003,HowesJGR2008,SchekochihinAPJS2009,BoldyrevAPJ2013}. On the other hand, at $\betai=5$, all the spectra are steeper, fitted by a $-3$ slope. Steepening of the spectra are possibly due to features not included in the standard theories, such as compressibility and/or wave damping effects~\citep[][; see also Fig.~\ref{fig:Damping_KAW-whistler}]{AlexandrovaAPJ2008,HowesJGR2008}.
In the electric energy, $E_E$, at $\betai=5$ and $0.2$, a power law steeper than $-5/3$ is seen at $k_\perp\di<1$, whereas a spectral index between $-2/3$ and $-1$ is observed at $k_\perp\di>1$.
At $\betai=1$, instead, a bump is present at $k_\perp\di=k_\perp\rhoi\sim1$ that makes the spectrum appear steeper, with a slope of $-1.8$ and it only partially agrees with the other two cases at $k_\perp\di\gtrsim4$ (Fig.~\ref{fig:Spectra_comparison}). 
In all the $\beta$ cases, the $E_E$ spectrum at $k_\perp\di<1$ is dominated by the MHD term, $E_{\rm MHD}=-\uv\times\Bv$, whereas at $k_\perp\di>1$ it is dominated by the Hall term, $E_{\rm Hall}=\Jv\times\Bv/n$ (cf. Eq.~(\ref{eq:HVM_Ohm})). The electron pressure term, $E_{\nabla\Pe}=-\bnabla\Pe/n$, is always found to be sub-dominant with respect to $E_{\rm Hall}$. 
Finally, the electric energy overcomes its magnetic counterpart at $k_\perp\di\sim2$, regardless of the $\rhoi$-scale position. 

We now investigate the nature of turbulent fluctuations in the different $\beta$ regimes. First, we compare the levels of magnetic and density spectra, $E_B$ and $C_0E_n$ (with $C_0=[\betai(1+\tau)/2][1+\betai(1+\tau)/2]$), a method to distinguish between KAW ($C_0E_n\simeq E_B$) and whistler ($C_0E_n\ll E_B$) turbulence~\citep{ChenPRL2013}. Second, we check if the relation $C_1E_n\simeq E_{B\|}$ (with $C_1=[\betai(1+\tau)/2]^2$) between the density and parallel magnetic spectra, expected for KAW fluctuations, is satisfied~\citep{ChenPRL2013,SchekochihinAPJS2009,BoldyrevAPJ2013}. We stress that the two methods are not conclusive if taken separately, but are complementary to each other and must therefore be inspected accordingly.

In Fig.~\ref{fig:Spectra_KAW-or-whistler_A} we compare the magnetic and the normalized density spectra, $E_B$ and $C_0E_n$, as obtained in our simulations. The main result is that the turbulence is mediated by magnetosonic/whistler (MS/W) fluctuations at $\betai=0.2$, whereas the dynamics at $\betai=1$ appear to be dominated by Alfv\'en wave/kinetic Alfv\'en wave (AW/KAW) turbulence. At $\betai=5$, instead, there is a signature of a transition at $k_\perp\di\sim1$, from a MS to a KAW regime. In order to confirm this scenario, in Fig.~\ref{fig:Spectra_KAW-or-whistler_B} we show the comparison between $E_{B\|}$ and $C_1E_n$. In particular, at $\betai=0.2$, a significant disagreement between the two quantities remains even at $k_\perp\di>10$, thus providing a confirmation of the whistler-dominated regime inferred from Fig.~\ref{fig:Spectra_KAW-or-whistler_A}. At $\betai=1$, we find $C_1E_n\simeq E_{B\|}$ through the entire $k_\perp$ range, thus confirming the KAW-dominated scenario. At $\betai=5$, $E_{B\|}$ and $C_1E_n$ differ by more than an order of magnitude for $k_\perp\di\lesssim1$, whereas the relation $C_1E_n\simeq E_{B\|}$ holds well for $k_\perp\di>1$. This supports the interpretation of a transition from an MS dynamics at large scales to a KAW regime at smaller scales, for $\betai=5$ (see also Sec.~\ref{subsec:interpret} and Fig.~\ref{fig:Damping_KAW-whistler}). Further evidence leading to the above conclusions is provided by inspecting the magnetic compressibility, $C_\|\equiv\delta B_\|^2/\delta B^2$, and by the predominantly perpendicular heating
of the ions at low $\beta$, $T_\perp>T_\|$ (not shown here).

%================================
%================================
%================================
%================================
We finally caution that these results may be dependent on the details of how the turbulence is driven. While we concluded from several test simulations (not shown) that the qualitative results presented here do not depend significantly on the resolution and/or on the forcing amplitude, we also found that  MS/W waves are not excited in complementary test simulations conducted with a purely incompressible (perhaps somewhat idealized) driving ($\alpha_1=1$, $\alpha_2=0$). Studying the detailed dependence of this kind of turbulence on the driving lies outside the scope of the present paper, but may also be relevant to the solar wind context and will therefore be worth exploring in the future.\\

\subsection{A possible interpretation}\label{subsec:interpret}

There are several examples of wave-supporting turbulent systems where linear physics leaves an imprint on the nonlinear dynamics even in strong turbulence regimes~\citep{NazarenkoJFM2011,TenBargeApJ2012,ChenPRL2013,KiyaniApJ2013,HadidApJ2015}. While it may not apply quantitatively in such regimes, linear theory may still provide some interesting physical insights into the dynamics at work in that case.
%================================
\begin{figure}[!b]
 \begin{minipage}[!h]{0.325\textwidth}
  \flushleft\includegraphics[width=1.15\textwidth]{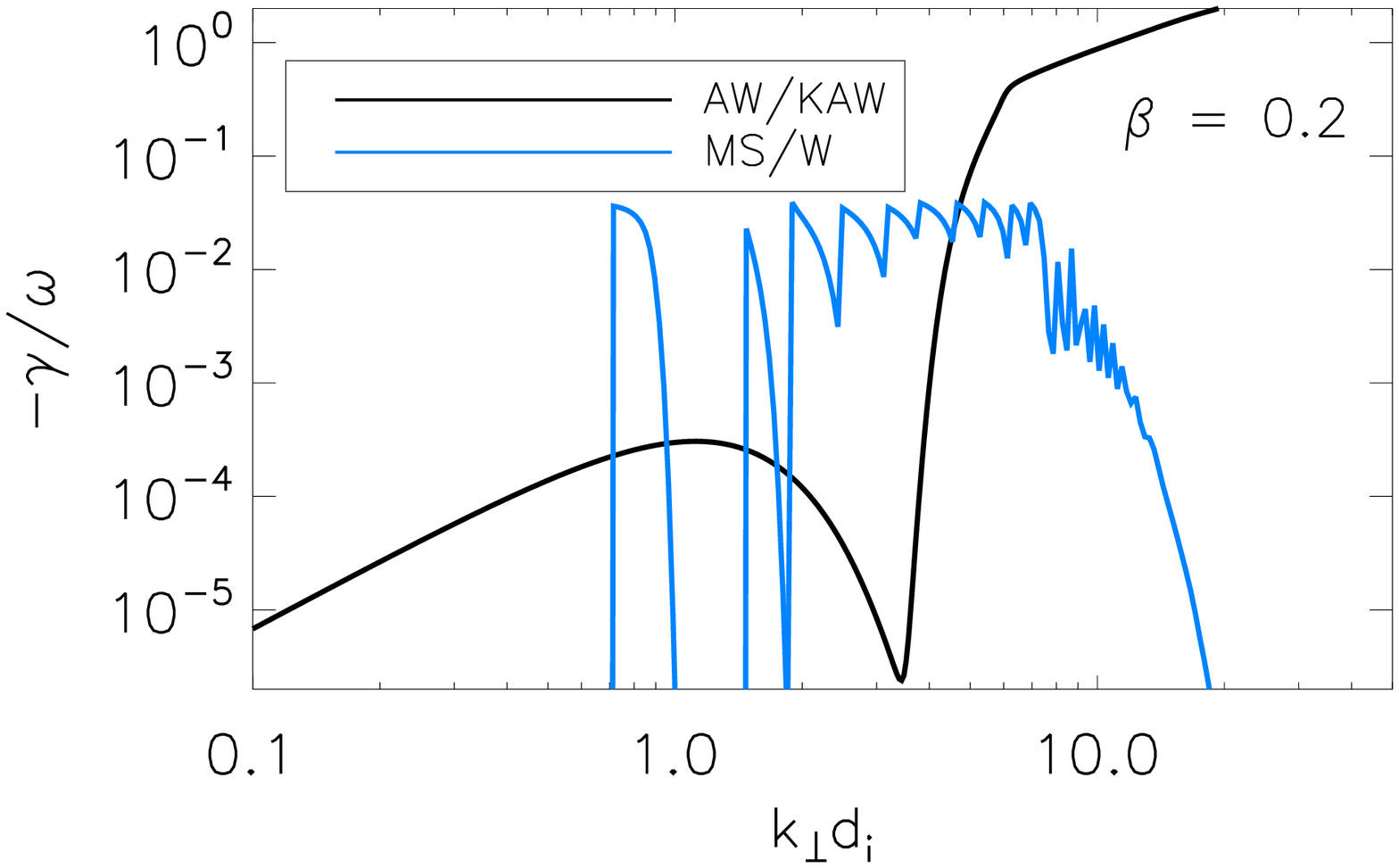}
 \end{minipage}
 \begin{minipage}[!h]{0.325\textwidth}
  \flushleft\includegraphics[width=1.15\textwidth]{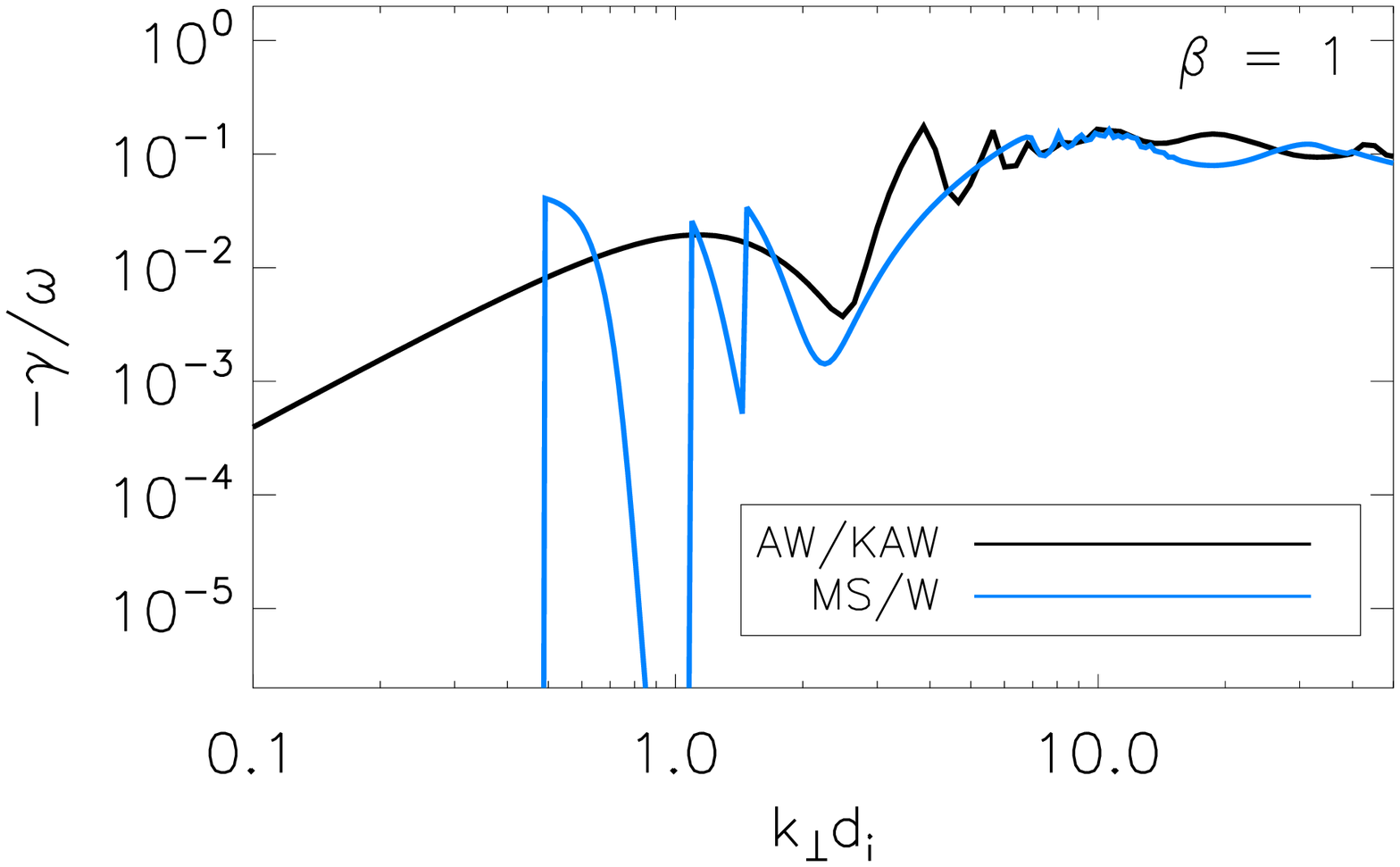}
 \end{minipage}
 \begin{minipage}[!h]{0.325\textwidth}
  \flushleft\includegraphics[width=1.15\textwidth]{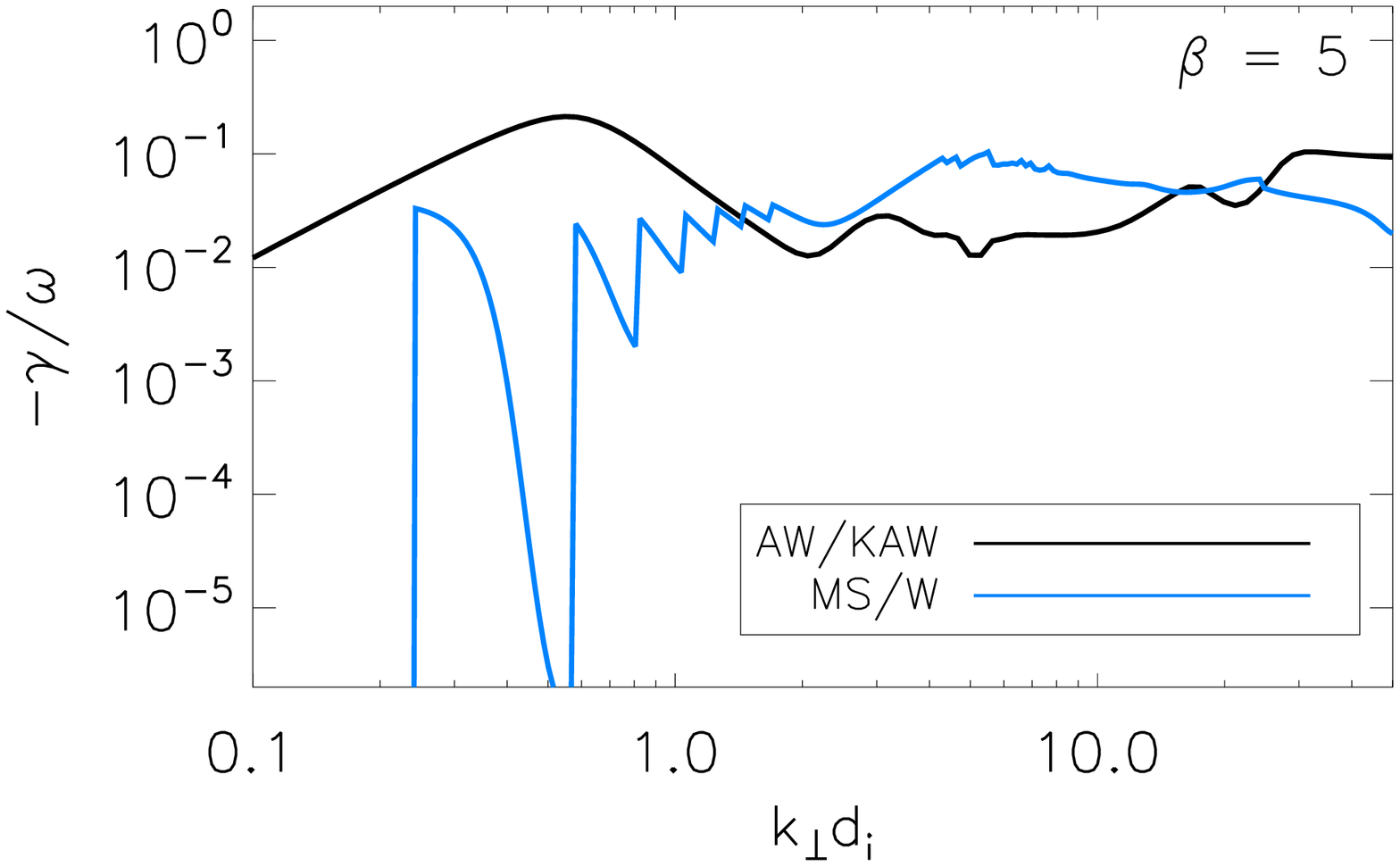}
 \end{minipage}
 \caption{Damping rate normalized to the linear wave frequency, $-\gamma/\omega$, for the AW/KAW (black) and MS/W (cyan) branches at $\beta=0.2$ (left panel) and $\beta=1$ (central panel), for a propagation angle of $\vartheta\simeq85^\circ$, and at $\beta=5$ (right panel), for $\vartheta\simeq86^{\circ}$.}
 \label{fig:Damping_KAW-whistler}
\end{figure} 
%================================ 

A possible interpretation for the transition reported above is in terms of the linear properties of the magnetosonic/whistler (MS/W) and of the Alfv\'en/kinetic Alfv\'en (AW/KAW) modes. In Fig.~\ref{fig:Damping_KAW-whistler} we display the ratio of the damping rate to the real frequency, $-\gamma/\omega$, for the AW/KAW and the MS/W branches of the HVM system, Eqs.~(1)-(3), within our simulation parameters. A representative propagation angle of $\vartheta\sim85^\circ$ has been estimated by $k_\|/k_\perp\sim\langle B_\perp/B\rangle$.
 
At $\betai=0.2$, the AW/KAW is weakly damped for $k_\perp\di\lesssim3$ and undergoes a complete resonant absorption as $\omega\to\Omegaci$ for $k_\perp\di>3$. The MS/W mode is instead practically undamped for $k_\perp\di\lesssim3$, except for a well-separated series of peaks representing the crossing of the resonant surfaces $\omega-n\,\Omegaci=0$ ($n=1$, $2$, $3$, $\dots$). Then, for $k_\perp\di>3$, the peaks form a quasi-continuum of wave damping, but still more than one order of magnitude lower than that of the AW/KAW counterpart. This would suggest a complete absorption of KAWs for $k_\perp\rhoi>1$ at $\betai=0.2$, leaving this regime whistler-dominated. 

At $\betai=1$, the frequency-normalized damping rates of the two modes are comparable and an extrapolation to the turbulent state is not obvious (a comparison of just $-\gamma$ would show a slightly higher damping of the MS/W branch, but still of the same order of magnitude). However, in this regime the AW/KAW mode is not completely absorbed anymore by the ion cyclotron resonance, consistent with a KAW-dominated cascade inferred from the simulations.

At $\betai=5$, instead, the frequency-normalized damping rates exhibit a transition at $k\di\sim1$: for $k\di\lesssim1$, the AW/KAW branch is more damped than the MS/W counterpart, whereas at $k\di>1$ the contrary holds (this transition is much more pronounced in the pure damping rates, $-\gamma$). This reflects the behavior shown in Fig.~2--3 at $\betai=5$, from which a transition from an MS-regime at $k\di<1$ to a KAW-dominated scenario for $k\di>1$ was inferred.

We point out that the electron damping on both the AW/KAW and the MS/W modes is missing in the HVM system. This represents a limitation of this model, which should be properly investigated as appropriate numerical resources become available. Nevertheless, we note that the interpretation proposed above is in qualitative agreement with previous linear studies in a full-kinetic framework~\citep{GaryJGR2009} and with observations about the relevance of cyclotron-resonant dissipation mechanisms in some regimes of SW turbulence~\citep{BrunoAPJL2014,BrunoAPJL2015}.

\section{Conclusions}

We presented the first high-resolution simulations of 2D3V forced hybrid-kinetic turbulence ranging from magnetohydrodynamic scales to scales well below the ion gyroradius. The spectral properties of the simulated turbulence, such as power-law exponents and spectral breaks at ion scales, are in agreement with the existing theory of subproton-scale turbulence and close to the observed SW
spectra. Moreover, we find that small-scale turbulence in this driven 2D3V setup mainly involves magnetosonic/whistler fluctuations at low $\beta$, and KAWs at somewhat higher $\beta$. We found that this transition correlates with a change in the relative strength of the damping of the underlying wave modes, suggesting that cyclotron-resonant damping may be relevant in this context. We point out that this scenario is not mutually exclusive of other important effects involving nonlinearities, such as the presence of coherent structures also spotted in the simulations, and they can in fact be coupled with each other. 

While the model used in this paper presents some limitations and does not accommodate all the dynamical complexity of the SW, the results suggest a possible dependence of subproton-scale kinetic turbulence on the plasma $\beta$ parameter that may be relevant to the time and space variability of the SW. High-resolution simulations in three spatial  dimensions, also including electron kinetic effects and different forms of driving, appear necessary to further our understanding of this problem, but will have to wait until computational capabilities become available.

%*******************%
%   Acknowledments  %
%*******************%
\acknowledgments

The authors acknowledge useful discussions with J.~M.~TenBarge,
A.~A.~Schekochihin, W.~Dorland, M.~Kunz, R.~Bruno and
F.~Pegoraro. We gratefully acknowledge the anonymous referee,
  whose in-depth comments helped to significantly improve the
  presentation and discussion of the results.
The research leading to these results has received funding from the
European Research Council under the European Union’s Seventh Framework
Programme (FP7/2007-2013)/ERC Grant Agreement No. 277870. This project
has received funding from the Euratom research and training programme
2014-2018. This work was facilitated by the Max-Planck/Princeton
Center for Plasma Physics. The simulations were performed on Fermi
(CINECA, Italy) and on Hydra (Rechenzentrum Garching, Germany).\\ \bigskip

%******************%
%   Bibliography   %
%******************%
\bibliographystyle{apj}

\end{document}